# Quantum-Noise-Limited Sensitivity-Enhancement of a Passive Optical Cavity by a Fast-Light Medium

David D. Smith[a], H. A. Luckay[b], Hongrok Chang[c], Krishna Myneni[d],

[a]Space Systems Department, NASA Marshall Space Flight Center, ES34, Huntsville, AL 35812

[b]Torch Technologies, 4035 Chris Dr. Suite C, Huntsville, AL 35802

[c]General Atomics, Electromagnetic Systems, 678 Discovery Dr., Huntsville, AL 35806

[d]U.S. Army Aviation & Missile Research Development and Engineering Center, RDMR-WDS-MRI, Redstone Arsenal, AL 35898

## ABSTRACT

We demonstrate for a passive optical cavity containing a dispersive atomic medium, the increase in scale factor near the critical anomalous dispersion is not cancelled by mode broadening or attenuation, resulting in an overall increase in the predicted quantum-noise-limited sensitivity. Enhancements of over two orders of magnitude are measured in the scale factor, which translates to greater than an order-of-magnitude enhancement in the predicted quantum-noise-limited measurement precision, by temperature tuning a low-pressure vapor of non-interacting atoms in a low-finesse cavity close to the critical anomalous dispersion condition. The predicted enhancement in sensitivity is confirmed through Monte-Carlo numerical simulations.

**Keywords:** Optical Resonators, Laser Gyroscopes, Coherent Optical Effects, Anomalous Dispersion, Fast Light

## 1. INTRODUCTION

The suggestion that the sensitivity of a passive optical cavity to a change in optical path length can be increased by placing an absorbing medium inside the cavity is counterintuitive. Near their resonance frequencies, not only do the atoms absorb energy decreasing signal-to-noise, the cavity line-width also broadens as a result of the associated anomalous dispersion, i.e., the white-light cavity effect. The combined effect would presumably reduce measurement sensitivity. However, this does not take into account the increased shift in the cavity mode frequency in response to a change in cavity length (or other external stimulus), i.e., scale-factor enhancement, which also occurs in the vicinity of such an absorption resonance. Recent publications have demonstrated that the scale factor is enhanced provided the dispersion is anomalous. Furthermore, the scale factor increases faster than the cavity line-width as the absorption and corresponding anomalous dispersion are increased, provided the cavity is under-coupled, resulting in a net increase in the scale-factor-to-mode-width ratio [1-3]. These increases can be particularly dramatic as the cavity approaches a critical anomalous dispersion (CAD) condition. Nevertheless, the increase in the scale-factor-to-mode-width ratio is accompanied by a substantial decrease in signal-to-noise as a result of medium absorption [4]. The question of whether the overall measurement sensitivity of a passive cavity can be increased has, therefore, persisted [5].

We address this question by simultaneously measuring the change in scale factor, mode width, and signal-to-noise as a low-finesse cavity is scanned in optical path-length and one of its modes passes through the $D_2$ $F_g = 2$ to $F_e$ Doppler-broadened resonance of a low-pressure vapor of non-interacting $^{87}$Rb atoms. By temperature tuning the atomic vapor, a pole in scale factor is observed near the CAD condition. A semi-empirical absorption/dispersion model [6], valid at low intensities, is used to extract the temperatures and scale-factor enhancements, while the change in mode-width and signal-to-noise are directly measured. The change in frequency resolution for the dispersive cavity in comparison with the corresponding empty cavity can then be inferred from these measured quantities. To check our assumptions and compute the frequency uncertainties directly, rather than through inference, we perform Monte-Carlo simulations using the cavity spectrum determined from the theoretical model as a master probability distribution function, from which sample photon distributions are randomly drawn. Finally, we determine whether an enhancement in sensitivity to changes in optical path length can occur under ideal (quantum-noise-limited) circumstances, i.e., whether the increase in scale factor can be larger than the associated increase in frequency uncertainty.

The low-finesse of the cavity is particularly important in the experiments and simulations described above because it ensures: (*i*) the intracavity intensity is low enough for application of the absorption model, (*ii*) the cavity is not so under-coupled that the attenuation of the modes causes them to disappear entirely when they are tuned across the absorption resonance, and (*iii*) the width of the mode is sufficiently large that the Doppler-broadened absorption profile is not effectively uniform across it, i.e., the dispersion cannot be approximated as linear over the finite width of the mode. Whereas enhancement of the scale factor only requires linear dispersion, enhancement in the scale-factor-to-mode-width ratio requires higher-order dispersion. The latter results in mode reshaping which limits mode broadening and further enhances the scale factor, resulting in an increase in the scale-factor-to-mode-width ratio. Without this mode reshaping, the scale factor and mode width would increase by the same amount and their ratio would not increase above unity. Mode reshaping is, therefore, crucial to achieving an increase in sensitivity. The relevant parameter to obtain mode reshaping is the ratio of the cavity-mode-width to absorption resonance width, which should not be too much smaller than unity. Therefore, high-finesse cavities can also be enhanced but require narrower absorption features.

Previously, we measured the cavity scale-factor enhancement as the temperature of the intracavity dispersive atomic medium was tuned, and demonstrated how temperature fluctuations in the atomic medium limit the scale-factor stability [6] and, therefore, the measurement precision. In the present work, we limit our discussion to the quantum noise limit (QNL) by ignoring classical noise, and therefore neglect the effect of these temperature fluctuations as they are not relevant in the QNL. Furthermore, to address the question at hand, we extend the previously-obtained experimental results, which were limited to scale factor, by providing additional measurements of the cavity mode width and signal-to-noise while temperature tuning.

An enhancement in optical cavity sensitivity using anomalous dispersion could have applications ranging from detection of gravity waves [7-10], tests of general relativity [11], optical communications schemes [12-14], enhanced strain and displacement sensing [15], wideband coherent perfect absorbers [3, 16], and increasing the precision of optical gyroscopes [1-3, 6, 11, 17-20]. Moreover, the fast-light enhancement is a particular example of a broader phenomenon that can be found in any physical system that possesses an exceptional point [3]. Such points are commonly found in non-Hermitian systems such as coupled oscillators having different loss rates. The critical anomalous dispersion is, in fact, an exceptional point, arising from the coupling of atomic and cavity resonant modes. Therefore, the enhancement in sensitivity described herein could ultimately benefit applications that extend far beyond those that rely on optical-cavity-based sensing.

In section 2 of the text, the enhancement in the quantum-noise-limited (QNL) precision of a dispersive cavity is derived in terms of the scale factor, mode width, and signal-to-noise. In section 3 the scale-factor enhancement in transmission and in reflection is derived for a cavity containing an atomic medium, and in section 4 an applicable model [6] for the absorption and phase shift of the medium is presented. Sections 5 and 6 discuss the experimental results, while section 7 describes the results of the Monte-Carlo simulations.

## 2. QUANTUM-NOISE-LIMITED PRECISION OF A DISPERSIVE CAVITY

A useful metric for a sensor utilizing the measurement of cavity mode frequency displacement for a given change in cavity optical path length, e.g., in an optical ring cavity gyroscope, is given by

$$\chi = \frac{\omega_q - \omega'_q}{\delta\omega}, \tag{1}$$

where $\omega_q$ is the peak frequency of the $q^{\text{th}}$ cavity mode, $\omega'_q$ is its peak frequency at the initial or reference cavity length, and $\delta\omega$ is the measurement uncertainty in the peak frequency. Thus, if mode peak displacement $\omega_q - \omega'_q$ can be increased without a concomitant increase in $\delta\omega$, the sensitivity of the sensor is enhanced. The minimum resolvable input to the cavity may then be found from Eq. (1) by setting $\chi = 1$.

It follows that the sensitivity enhancement (or reduction) of a dispersive cavity, compared to the corresponding empty cavity, can be written as

$$\zeta \equiv \frac{\chi}{\chi^{(e)}} = \frac{\omega_q - \omega'_q}{\omega_q^{(e)} - \omega_q'^{(e)}} \cdot \frac{\delta\omega^{(e)}}{\delta\omega}. \tag{2}$$

We denote the mode frequency, initial mode frequency, measurement uncertainty, full-width-at-half-maximum (FWHM) linewidth, and signal-to-noise ratio of the empty cavity by $\omega_q^{(e)}$, $\omega_q'^{(e)}$, $\delta\omega^{(e)}$, $\gamma^{(e)}$, and $SNR^{(e)}$, respectively. The corresponding quantities for the dispersive cavity will not be superscripted by (*e*). The cavity scale-factor enhancement can then be defined as $S(\omega_q) \equiv d\omega_q / d\omega_q^{(e)}$. Hence, for a dispersion-enhanced cavity operating in the vicinity of an absorption resonance, a small change in the cavity's optical path-length will result in a mode peak displacement

$$\omega_q - \omega_q' = S\left(\omega_q^{(e)} - \omega_q'^{(e)}\right). \tag{3}$$

Therefore, regardless of the source of the uncertainties, the sensitivity enhancement can be written

$$\zeta = S \cdot \frac{\delta\omega^{(e)}}{\delta\omega}. \tag{4}$$

An equally valid way to arrive at Eq. (4) is to set $\chi = 1$ and realize that an error, $\delta\omega$, in a dispersive-cavity is equivalent to $\delta\omega / S$ in the corresponding empty-cavity. Note that the reduction in the minimum measurable cavity mode frequency is simply given by $1/\zeta$.

Now let us assume that all classical sources of error in the cavity can be eliminated, e.g., due to temperature and mechanical fluctuations, leaving only quantum-mechanical photon shot noise. It has been reported [21, 22] that the frequency error for an empty (unidirectional) passive cavity in this QNL is

$$\delta\omega_{QNL}^{(e)} \approx \frac{\gamma^{(e)}}{SNR^{(e)}}. \tag{5}$$

If we assume Eq. (5) also applies to a cavity containing a dispersive medium, then $\delta\omega_{QNL} \approx \gamma / SNR$. Accordingly, Eq. (4) becomes

$$\zeta^{QNL} \approx \frac{S \cdot M}{W}, \tag{6}$$

where $M = SNR / SNR^{(e)}$ is the normalized signal-to-noise and $W = \gamma / \gamma^{(e)}$ is the normalized cavity linewidth. Hence, if $\omega_q$ could be measured to the same QNL precision as $\omega_q^{(e)}$, the enhancement would simply be $\zeta^{QNL} = S$. However, because of mode broadening and attenuation, $\delta\omega_{QNL}$ is typically larger than $\delta\omega_{QNL}^{(e)}$. As we shall see, however, it is not larger by a factor of $S$.

Mode broadening and attenuation reduce, but do not cancel out, the increase in scale factor, resulting in an overall increase in the QNL sensitivity. Note that if, in Eq. (6), we take $SNR = SNR^{(e)}$, i.e., $M = 1$, then we obtain the sensitivity enhancement factor that we used in our earlier work [2], namely $\zeta^{QNL} = S/W$. Also, note that while each of the quantities in Eq. (6) are frequency dependent, we will primarily be concerned with their values at a particular frequency $\omega_0$, defined as the frequency where the scale factor enhancement is maximized (near the absorption peak). The QNL enhancement, scale-factor enhancement, normalized signal-to-noise, and normalized mode width at $\omega_0$ will be denoted by $\zeta_0^{QNL}$, $S_0$, $M_0$, and $W_0$, respectively.

In the experiment outlined in section 5, we use Eq. (6) to evaluate the peak QNL sensitivity enhancement, $\zeta_0^{QNL}$, by measurement of $S_0$, $M_0$, and $W_0$. The advantage of this approach is that one can predict the QNL enhancement without having to directly measure frequency errors, i.e., perform a full noise analysis. However, a critical assumption behind Eq. (6) is that Eq. (5) should apply to the dispersive cavity in the same proportion that it applies to the empty cavity. This assumption neglects the effect on the uncertainty of the change in mode shape that occurs as the CAD condition is approached (the mode becomes more flat-topped). In section 7, on the other hand, the assumptions leading to Eqs. (5) and (6), are abandoned, and we take Eq. (4) as our starting point for finding the QNL enhancement. The frequency errors $\delta\omega_{QNL}$ are calculated numerically via Monte-Carlo simulations using the cavity transmission spectrum, see Eq. (10), as a master photon probability distribution function, from which sample photon distributions are drawn. Because they are drawn randomly, the statistics are Poissonian, and therefore the computed frequency errors are the QNL quantities of interest. The importance of this numerical procedure is that it bypasses the assumptions behind Eq. (6), providing a more direct indication of the QNL enhancement that takes full account of the change in mode shape near the CAD condition. We find that the RHS of Eq. (6) must be modified by a constant factor *K* that depends on the particular form of averaging employed in finding the distribution peak.

We assume that the source of the quantum noise in our experiments is detector shot noise and any additional sources of quantum noise, e.g., due to spontaneous emission in the medium, are negligible. This is a good assumption for passive cavities because medium absorption and optical pumping are much weaker, and therefore the relative amount of spontaneous emission to laser light is much smaller, than for the case of a laser (where spontaneous emission noise cannot be neglected). The signal-to-noise for shot-noise-limited direct detection of photons on a photodiode is

$$SNR \sim \frac{N_S}{\sqrt{N_S}} = \sqrt{N_S}, \qquad (7)$$

where $N_S$ is the number of signal photons per second arriving at the detector. In this case the signal is associated with the resonance of a cavity. Eq. (7) can be generalized to account for cavity fringe baselines by substituting $N_S - N_B$ for $N_S$ in the numerator and $N_S + N_B$ for $N_S$ (variances add) in the denominator of Eq. (7), where $N_B$ is the number of photons per second arriving at the detector for the baseline (away from the cavity resonance). This is equivalent to substituting $(N_S - N_B)^2 / (N_S + N_B)$ for $N_S$ on the RHS of Eq. (7). The signal-to-noise can, therefore, be obtained from transmission or reflection spectra by using the relation

$$\frac{(N_S - N_B)^2}{N_S + N_B} = \left(\frac{m^2}{2 \pm m}\right)\frac{P_B}{\hbar\omega} = \left(\frac{2V^2}{1 \mp V}\right)\frac{P_B}{\hbar\omega}, \qquad (8)$$

where $m = |N_S - N_B| / N_B$ is the fringe modulation depth, *V* is the fringe visibility, and $P_B$ is the baseline power, and the positive (negative) sign is used in transmission (reflection). Note that the factor $P_B / \hbar\omega$ is the same for the dispersive and empty cavities and hence factors out in the determination of the ratio, *M*. Therefore,

$$M = \frac{m}{m^{(e)}} \sqrt{\frac{2 \pm m^{(e)}}{2 \pm m}} = \frac{V}{V^{(e)}} \sqrt{\frac{1 \mp V^{(e)}}{1 \mp V}}, \qquad (9)$$

where the relation $m = 2V/(1 \mp V)$ has been used to transform to fringe visibility. Again, the superscript (*e*) is used to denote the empty-cavity fringe modulation depth and visibility.

## 3. SCALE FACTOR OF A DISPERSIVE CAVITY

The complex reflection and transmission coefficients for a Fabry-Pérot or two-port ring cavity containing a dispersive atomic medium at temperature *T* can be written as

$$\tilde{\rho}(\omega,T) = \frac{1}{r_1} \frac{r_1^2 - \tilde{g}(\omega,T)e^{i\omega\tau_c}}{1 - \tilde{g}(\omega,T)e^{i\omega\tau_c}}, \qquad (10)$$

and

$$\tilde{\tau}(\omega,T) = \frac{t_1 t_2}{\sqrt{r_1 r_2}} \frac{\sqrt{\tilde{g}(\omega,T)}e^{i\omega\tau_c/2}}{1 - \tilde{g}(\omega,T)e^{i\omega\tau_c}}, \qquad (11)$$

respectively, where $\tau_c$ is the round-trip time of the empty cavity. The net electric-field gain per round trip is $\tilde{g}(\omega,T) = r_1 r_2 a_1 \tilde{\tau}_m(\omega,T)$, where $r_{1,2}$ and $t_{1,2}$ are the real-valued reflection and transmission coefficients of the mirrors, respectively, $a_1$ accounts for other frequency-independent round-trip losses in the cavity, and $\tilde{\tau}_m(\omega,T) = \tau_m(\omega,T)\exp[i\Phi(\omega,T)]$ is the complex round-trip transmission coefficient of the atomic medium.

The resonance condition is then determined by finding the frequencies $\omega_q$ where the derivative of the transmission $|\tilde{\tau}(\omega,T)|^2$ or reflection $|\tilde{\rho}(\omega,T)|^2$ with respect to $\omega$ goes to zero, i.e., from the transcendental equation

$$\omega_q \tau_c + \Phi(\omega_q,T) + F(\omega_q,T) = 2\pi q = \omega_q^{(e)} \tau_c, \qquad (12)$$

where *q* is the cavity mode number, $\Phi(\omega,T)$ is the effective round-trip phase-shift of the medium, and $F(\omega,T)$ is an additional phase factor that arises from the reshaping of the mode by the medium absorption. The scale-factor enhancement is then obtained by taking the derivative of the dispersive cavity mode frequencies with respect to the empty cavity mode frequencies,

$$S(\omega_q,T) \equiv \frac{d\omega_q}{d\omega_q^{(e)}} = \frac{1}{\hat{n}_g(\omega_q,T) + T_{cav}(\omega_q,T)}, \qquad (13)$$

where $\hat{n}_g(\omega,T) = 1 + (1/\tau_c) d\Phi/d\omega$ is the effective group index and $T_{cav}(\omega,T) = (1/\tau_c) dF/d\omega$ is the additional dimensionless time delay associated with absorption. Note that the scale factor has a pole at a critical anomalous dispersion, and is enhanced when the sum of the group index and this additional cavity delay time is less than unity, i.e., the sum of the slopes or net time delay contributed by these two terms must be negative for an enhancement to occur.

The functional form of $F$ in transmission differs from that in reflection, which leads to different scale factors for transmission and reflection. In transmission

$$F(\omega,T) = -\sin^{-1}\left[\frac{1 - g(\omega,T)^2}{2g(\omega,T)^2} \frac{g'(\omega,T)}{\hat{n}_g(\omega,T)\tau_c}\right], \qquad (14)$$

where $g'(\omega,T) = dg/d\omega$. On the other hand, in reflection

$$F(\omega,T) = -\sin^{-1}\left[\frac{A(\omega,T)}{\sqrt{A^2(\omega,T) + B^2(\omega,T)}}\right] + \sin^{-1}\left[\frac{C(\omega,T)}{\sqrt{A^2(\omega,T) + B^2(\omega,T)}}\right], \qquad (15)$$

where

$$\begin{aligned}
A(\omega,T) &= g'(\omega,T)\left(r_1^2 + g(\omega,T)^2\right) \\
B(\omega,T) &= g(\omega,T)\hat{n}_g(\omega,T)\tau_c\left(r_1^2 - g(\omega,T)^2\right) \\
C(\omega,T) &= g'(\omega,T)g(\omega,T)\left(r_1^2 + 1\right).
\end{aligned} \qquad (16)$$

Note that while the phase shift $\Phi(\omega_q,T)$ influences the mode frequencies by mode pushing, $F(\omega_q,T)$ does so instead by mode reshaping. When the cavity mode is narrow in comparison with the atomic resonance, i.e., in the high-finesse (linear-dispersion) approximation, the absorption variation across the mode and resultant mode reshaping can be neglected, and we obtain $S(T) = 1/\hat{n}_g(T)$ [11, 20]. The cavity linewidth is given, within this approximation, by $\gamma(T) = (4/\hat{n}_g(T)\tau_c)\sin^{-1}\sqrt{(1-g_0(T))/2(1+g_0(T))}$. Therefore, the normalized cavity linewidth can be written as

$$W(T) = \frac{\gamma(T)}{\gamma^{(e)}} = \frac{u(T)}{\hat{n}_g(T)} \tag{17}$$

where $u(T)$ takes into account the change in the resonant value of the net round-trip field gain, $g$, with temperature. The function $u(T)$ approaches unity at low temperatures and exceeds unity at all other temperatures. Therefore, $W(T)$ is always greater than $1/\hat{n}_g(T)$. Hence, if the linear-dispersion approximation were valid for all scale factors, there could be no enhancement of the scale-factor-to-mode-width ratio or QNL sensitivity, because broadening of the mode would always cancel the increase in scale factor. This cannot happen at all scale factors, however, because the broadening of the mode ensures that at some point the linear-dispersion approximation must be violated, in particular as the width of the mode approaches that of the medium resonance. Therefore, the scale factor and critical anomalous dispersion are more properly determined by Eq. (13). In addition, the mode width does not increase as fast as predicted by Eq. (17), but is instead clamped by the finite medium absorption width. The increased scale factor and decreased mode width, in comparison with the linear-dispersion prediction, work together to increase the scale-factor-to-mode-width ratio and QNL sensitivity.

According to Eq. (15) there is another way to obtain $F(\omega_q, T) = 0$, namely when $A(\omega, T) = C(\omega, T)$. The ratio $A(\omega, T)/C(\omega, T)$ thus determines whether mode reshaping from $F(\omega_q, T)$ will add to or subtract from the mode pushing due to $\Phi(\omega_q, T)$, i.e. whether the slope $T_{cav}(\omega, T)$ will be negative or positive, respectively. Enhancement of the scale-factor-to-mode-width ratio (and, therefore, of the QNL sensitivity) requires that the mode reshaping adds to the mode pushing, i.e., $T_{cav}(\omega, T) < 0$ or $A(\omega, T) > C(\omega, T)$. In reflection this only occurs for $g(\omega,T) < r_1^2$, or $r_2 a_1 \tau_m(\omega,T) < r_1$, i.e., when the cavity is under-coupled. This requirement is always fulfilled for symmetric cavities $(r_1 = r_2)$. Single-port (all-pass) cavities like the one presented in section 5, on the other hand, require $a_1 \tau_m(\omega,T) < r_1$. For over-coupled cavities the scale factor increase is more than offset by increased mode width, resulting in a decrease in the scale-factor-to-mode-width ratio. In transmission, on the other hand, only one term appears in Eq. (14), and the mode reshaping always adds to the mode pushing.

Note Eq. (10) assumes that any mirror losses, other than reflection, in the transmission through the input mirror are negligible, i.e., $r_1^2 + t_1^2 = 1$. We can take any additional loss into account by assuming the transmission coefficient of the first mirror is $t_1^2 = 1 - r_1^2 - b_1^2$, where $b_1$ accounts for this additional loss. The form of Eq. (11) and subsequent equations for transmission are unchanged by the added mirror loss. The equations for reflection, on the other hand, require substituting $g(1 - b_1^2)$ for $g$ in the numerator of Eq. (10) and wherever it, or its derivative, appears in Eq. (16).

## 4. ABSORPTION AND DISPERSION MODEL

For a given cavity, the functions $\Phi(\omega,T)$ and $F(\omega,T)$ are uniquely determined by the medium absorption coefficient $\alpha(\omega,T)$. As we discussed previously in detail [6], a semi-empirical model that takes into account all the Zeeman-degenerate Doppler-broadened hyperfine transitions of the $^{87}$Rb $D_2$ transition can be used to model the absorption provided the intensity is sufficiently weak that optical pumping, saturation, and power broadening are all negligible. Under these conditions the atoms remain in thermal equilibrium with level populations determined by the Boltzmann distribution. The absorption coefficient is then simply a weighted superposition of the various hyperfine transitions,

$$\alpha(\omega,T) = \alpha_D(T)\sum_{j=0}^{3} s(2,j)\exp\left[-\left(\frac{\omega-\omega_{2,j}}{\Gamma_D(T)}\right)^2\right], \quad (18)$$

where $\omega_{2,j}$ and $s(2,j)$ represent the frequencies and strengths of the various hyperfine transitions from the $F_g = 2$ ground state, respectively. The sum is over the excited states $F_e = j$. The temperature dependence of $\alpha(\omega,T)$ is determined by that of the 1/e Doppler width $\Gamma_D(T) = k\sqrt{2k_B T/m}$ and envelope function $\alpha_D(T) = C \cdot N(T)/\Gamma_D(T)$, where $C$ is constant to a good approximation. We assume the number density $N(T)$ follows the empirical relation $N(T) = (133.3/k_B T) \cdot 10^{(2.881+A+B/T)}$, where $A = 4.312$, $B = -4040$, and $T$ is specified in Kelvins [23]. Note, the coefficient $\alpha_D(T)$ represents the peak absorption coefficient only when the absorption can be properly described by a single Gaussian. Assuming linear polarization such that only transitions between states having the same magnetic quantum number are allowed, and ignoring Zeeman pumping so that the distribution of ground-state magnetic sublevels is uniform, the coefficients $s(2,j)$ for the allowed transitions are $s(2,1) = 0.05$, $s(2,2) = 0.25$, $s(2,3) = 0.7$, and $C = (5/8)\sigma_0^\pi \sqrt{\pi}\,\Gamma/2 = 4.1\times 10^{-6}\, m^2/s$, where $\Gamma$ is spontaneous decay rate from the excited state and $\sigma_0^\pi = 1.938\times 10^{-13}\, m^2$ is the total on-resonant absorption cross-section for π-polarized light.

The medium effective phase-shift $\Phi(\omega,T)$ can then be determined from a Kramers-Krönig (K-K) relation by taking the Hilbert transform of the medium transmission coefficient $\tau_m(\omega,T) = \exp[-\alpha(\omega,T)\ell/2]$, where $\ell$ is the length of the atomic medium. For a Gaussian absorption profile this transform has the simple analytic form $H\{\exp[-a\exp(-x^2/b^2)]\} = ia\cdot\text{erf}(ix/b)\cdot\exp(-x^2/b^2)$. The phase shift is found by applying this transformation to each of the individual hyperfine transitions, i.e.,

$$\Phi(\omega,T) = \frac{i\alpha_D(T)\ell}{2}\sum_{j=0}^{3} s(2,j)\exp\left[-\left(\frac{\omega-\omega_{2,j}}{\Gamma_D(T)}\right)^2\right]\cdot\text{erf}\left(i\frac{\omega-\omega_{2,j}}{\Gamma_D(T)}\right). \quad (19)$$

## 5. DESCRIPTION OF EXPERIMENT

The experiment is the same as that we reported previously [6], except here we also measure mode widths and signal-to-noise, and our approach to the analysis of the results differs (see section 6). To summarize, a linearly-polarized external-cavity diode laser at a wavelength of 780 nm was used to scan over the modes of a 40 cm ring cavity, containing a 2.5 cm isotopically-enriched $^{87}$Rb quartz vapor-cell enclosed in a temperature-stabilized aluminum oven [6, 24]. An all-pass cavity configuration ($r_2 = 1$, $|\tilde{\tau}(\omega,T)|^2 = 0$) consisting of three high reflectance mirrors and a 90:10 cube beam-splitter for the input/output coupler was used, as shown in Fig. 1. The center frequency of the laser was adjusted to coincide with the Doppler-broadened $F_g = 2 \to F_e$ resonance and the laser was scanned over several cavity free spectral ranges. A Michelson interferometer with unequal arm lengths and a second $^{87}$Rb cell in a counter-propagating pump-probe saturated-absorption spectrometer were used as frequency references, and a reference detector was used to factor out variations in the laser intensity during the spectral scan. The incident beam was s-polarized, and the detuning between the cavity and atomic resonance was varied by an intracavity liquid-crystal variable retarder whose slow (tuning) axis was aligned with the polarization.

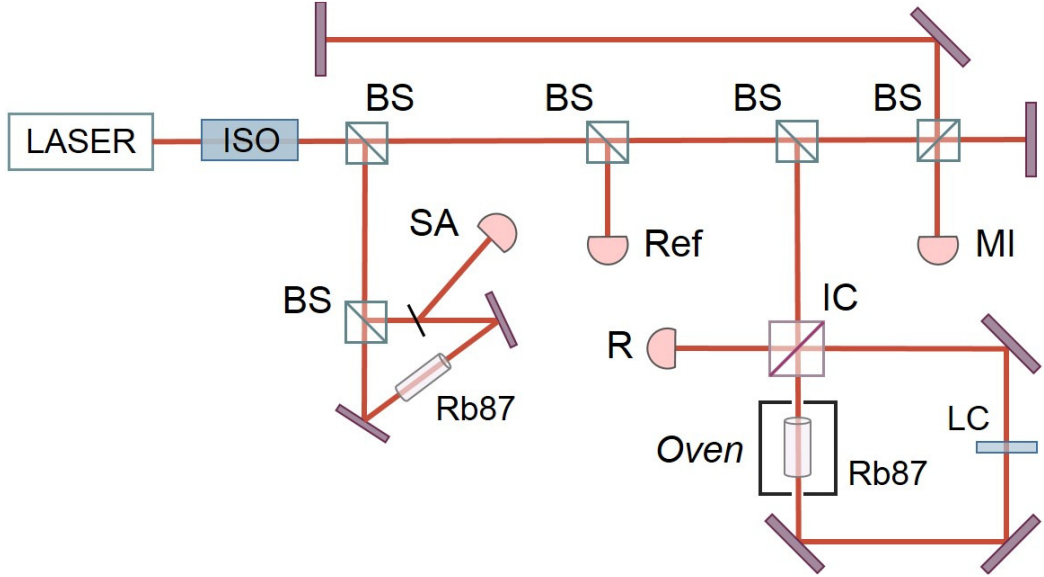

Fig. 1 (Color online) Experimental Setup. BS = Beam-Splitter, ISO = Isolator, LC = Liquid Crystal Variable Retarder, IC = Input Coupler, SA = Saturated Absorption Spectrometer, Ref = Reference Detector, MI = Michelson Interferometer, R = Cavity Reflection.

The temperature of the atomic vapor cell was varied from 36°C to 42.5°C through the critical anomalous dispersion (which occurred near 42°C). At each temperature, reflection spectra were recorded at a variety of detunings as a selected cavity mode was tuned across the atomic resonance via the variable retarder. Representative experimental spectra are shown in Fig. 2. A larger number of spectra were recorded close to the resonance by applying a nonlinear voltage step to the retarder. An automated peak finding program was used to obtain the frequencies, widths, and modulation depths of modes near to and far away from the resonance, resulting in plots of the mode detunings for the dispersive cavity vs. those for the empty cavity as shown in Fig. 4(a). In this figure, each data point represents a different liquid crystal voltage and the detunings of the dispersive and empty cavities are converted into linear FSR units, i.e., the detunings are defined as $\Delta = (\omega_q - \omega_0)\tau_c / 2\pi$ and $\delta = (\omega_q^{(e)} - \omega_0)\tau_c / 2\pi$, respectively, where $\omega_0$ is the frequency of the scale factor maximum, which, owing to the asymmetry of the absorption profile, is shifted slightly from the absorption peak. The subscript on the quantities $S_0$, $M_0$, $W_0$, and $\zeta_0^{QNL}$ refer to $\omega_0$, i.e., $S_0 \equiv S(\omega_q = \omega_0)$, etc.

## 6. EXPERIMENTAL RESULTS

In this section the relevant parameters $S_0$, $M_0$, and $W_0$ are obtained from the experimental reflection spectra for evaluation of $\zeta_0^{QNL}$ via Eq. (6). Representative spectra are shown in Fig. 2 for the case where the cavity is held near the CAD condition. Several points are worth remarking upon: (*i*) the effect of detuning the mode is greater for the mode nearest to the atomic resonance ($\Delta > \delta$), i.e., the scale factor is enhanced for this mode, (*ii*) this mode nearest resonance also broadens and is attenuated, (*iii*) the mode broadening is limited by the width of the atomic resonance, and (*iv*) the mode amplitude is never fully extinguished even at the CAD condition. Therefore, even without explicit evaluation of Eq. (6), simple observation of the cavity spectra suggests that a QNL sensitivity enhancement is possible, because at the CAD condition the scale factor enhancement approaches infinity, while the mode width and amplitude always remain finite and non-zero.

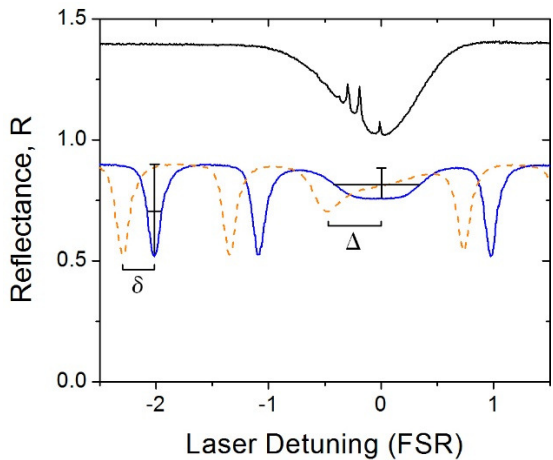

Fig. 2 (Color online) Representative reflection spectra data taken near the critical temperature (sensor temperature 314.25K, fit temperature 314.77K), at two different values of the empty-cavity detuning $\delta$ (solid

curve is for $\delta = 0$). A saturated absorption spectrum (top) is also shown for reference. As the mode approaches, it is pushed away and reshaped by the atomic resonance, resulting in a larger dispersive-cavity detuning $\Delta$. Consequently, the scale factor increases. The mode also broadens and is attenuated, but note the broadening is limited to the Doppler-broadened resonance width. As a result the attenuation is also less than expected. The on-resonance normalized mode width and signal-to-noise, $W_0$ and $M_0$, respectively, are calculated from measurements of the FWHM mode width and modulation depth (the dip depth divided by the fringe baseline), taken both on-resonance and away from the resonance, as shown.

The point of the current work is not to validate the theoretical model for $S_0$ vs. $T$ (as it was in our previous work [6]), but rather to use it to extract the maximum scale-factor enhancements by fitting to the experimental data. The following procedure was used to determine the scale-factor enhancements: For each spectrum, the FSR was measured. The average value of 1000 such measurements was 703.0 MHz with a standard deviation of 3.1 MHz, i.e., uncertainty of 0.1 MHz. The value of $\delta$ could not be directly measured, but was inferred by using a mode located sufficiently far from the influence of the atomic resonance and adding the appropriate number of FSR to its frequency. At each temperature, the two parameters, $r_1$ and $a_1$, were found by a nonlinear least-squares fit of the theory to the part of the reflection spectrum that was away from the atomic resonance. When input mirror losses (other than reflection) were ignored, the average values of these parameters were $a_1 = 0.674$ and $r_1 = 0.946$ with standard deviations of 0.044 and 0.006, respectively (over 22 measurements). When these losses were considered, the average best fit values were $a_1 = 0.672$, $r_1 = 0.917$ and $b_1 = 0.209$, with standard deviations of 0.020, 0.040, and 0.094, respectively (over 6 measurements). For comparison, direct measurement of the mirror transmission and reflection coefficients yielded $r_1 = 0.903$ and $b_1 = 0.260$. While the better agreement suggests that mirror losses should be considered, the inclusion of these losses ultimately makes no difference in the determination of $S_0$, $M_0$, or $W_0$, because the ratio of the cavity-mode-width to absorption resonance width is unchanged. From these parameter values it is obvious that the cavity was under-coupled, as required to obtain an enhancement in the scale-factor-to-mode-width ratio.

The largest value of the input intensity was 4.2 mW/cm² for these measurements (the typical value was 1.5 mW/cm²). However, it is the steady-state intracavity intensity that is important for determining the applicability of the absorption model. The average cavity finesse for 10 different temperature measurements was 7.6 with a standard deviation of 1.3. The intracavity intensity at the peak of a mode on-resonance with the atoms was always less than 20 µW/cm², far below the lowest saturation intensity for the hyperfine transitions with π-polarized light (3.6 mW/cm² for the $F_g = 2$ to $F_e = 3$ transition [23]). The low finesse and large on-resonance absorption self-limit the steady-state circulating intracavity intensity, enabling application of the model.

Each set of data in Fig. 4(a) consisted of 500 data points, obtained from spectra of the sort shown in Fig. 2. From each data set the scale-factor enhancement was determined by performing a nonlinear least-squares fit to the theory presented in sections 3 and 4 (Eq. (12) was the fit equation) with temperature as a fit parameter. The fitted temperatures are compared with those measured by a platinum RTD sensor mounted on the outside of the aluminum oven in Fig. 3. The strong correlation appears to confirm the theoretical model is valid for the intensities used in the experiment. However, systematic errors are likely present that weaken this argument. The most important of these are: (*i*) the sensor temperature may not accurately represent the temperature of the vapor (shifts the abscissa values), and (*ii*) the values of the *A* and *B* coefficients (see section 4) in the empirical relation for the number density may need to be adjusted (shifts the ordinate values), rather than relying on the published values [23] for an infinite liquid reservoir as we do here. We have observed that these coefficients vary over time for a given vapor cell. Estimated upper bounds for these systematic errors as determined from ancillary measurements [6] are +2K for the abscissa and +2.5K for the ordinate. What can be concluded, though, is that the optical detection/fitting procedure provides a more precise (if not necessarily accurate) temperature measurement than our temperature controller/sensor system. The estimated single-measurement temperature uncertainty of the RTD sensor due to temperature fluctuations and limited readout resolution was 150 mK (abscissa) (this value can be decreased through additional passive controls and/or by making multiple measurements), an order of magnitude larger than that obtained from the fitting procedure, 15 mK (ordinate).

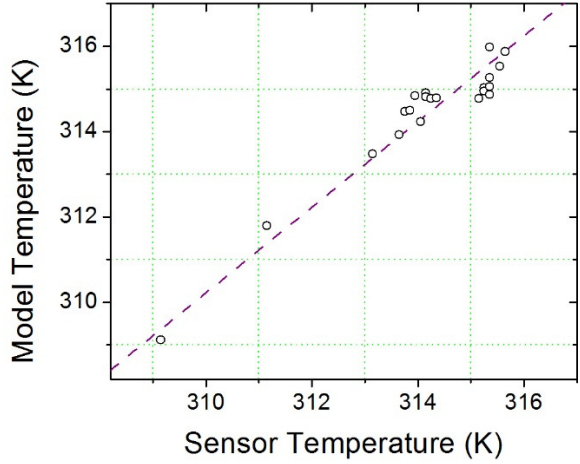

Fig. 3. (Color online) Temperature determined from the model vs. temperature measured by an oven-mounted sensor. The dashed line is a linear fit, weighted by the errors in both coordinates. The slope and intercept of this line were 1.004 and -0.926 K, respectively, with a correlation coefficient of 0.917.

The temperatures resulting from the fitting procedure and maximum on-resonance scale factors are plotted in Fig. 4(b). Note that a pole occurs in the scale factor corresponding to a critical anomalous dispersion of $\hat{n}_g = 0.483$ and critical temperature of $T_c = 314.82\,K$. This is far from the values predicted in the linear-dispersion approximation ($\hat{n}_g = 0$, $T_c = 322.13\,K$). The highest scale-factor enhancement recorded was $S_0 = 363$ (with errors of $+1175$ and $-157$). Note the error increases quadratically with the scale factor, and at some point will even exceed the value of the scale-factor enhancement itself. The error bars in Fig. 4(b) represent the uncertainties derived from the fit parameters assuming a 5 MHz error in the determination of $\Delta$ and a 6 MHz error in $\delta$. These errors primarily result from cavity noise during the spectral scan. (The FSR measurement makes the error in $\delta$ slightly larger than that of $\Delta$.)

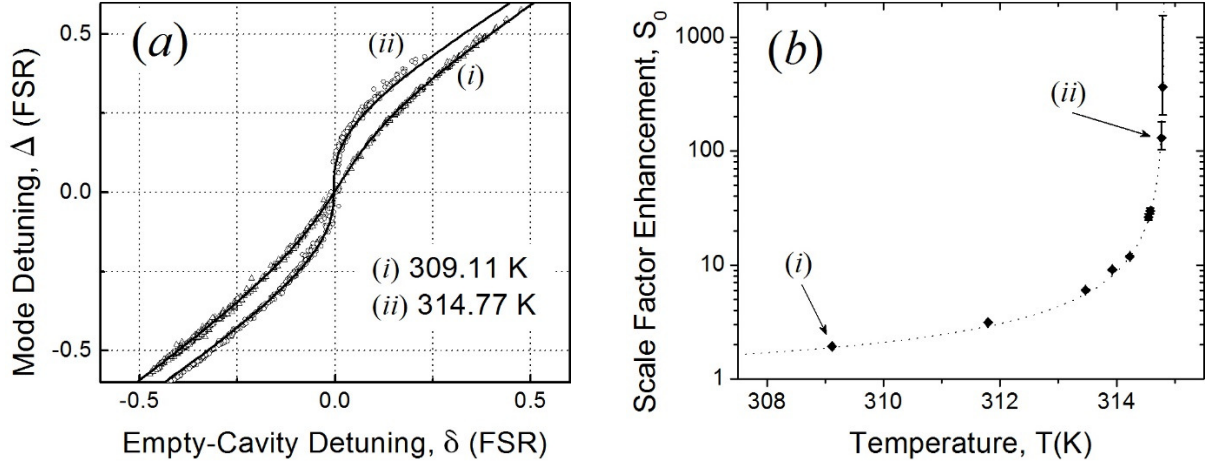

Fig. 4. (*a*) Scale factor plots near and far from the critical temperature (314.82 K). The temperatures are obtained from the nonlinear-least-squares fits (solid curves). The curves have been shifted slightly for clarity. (*b*) Scale-factor enhancement vs. the best fit temperature. The points (*i*) and (*ii*) represent the maximum scale factor slopes obtained from the corresponding data sets in (*a*). The dotted curve shows the expected results from the model when average values are used for $r_1$ and $a_1$.

From the cavity spectra, the on-resonance mode width and signal-to-noise were also measured and plotted normalized to their empty cavity (off-resonance) values in Fig. 5. The temperatures of the data points in Fig. 5(a) are those from Fig. 4(b), i.e., they were computed using the full theory. In Fig. 5(b), on the other hand, the temperatures were computed by using the linear-dispersion approximation to find the temperatures that correspond to the scale-factor enhancements in Fig. 4(b), leading to an average shift of +6.6K, slightly less than the shift in the critical temperature (+7.3K). The theoretical curves were determined numerically from spectra generated from the full theory (Fig. 5(a)), as well as analytically using the linear-dispersion approximation (Fig. 5(b)), i.e., Eq. (17), using the average values of $r_1$ and $a_1$. Note that for both theoretical curves, $M_0$ and $W_0$ approach unity at low temperatures, as expected when the medium does not absorb. At higher temperatures, however, the two theories diverge. In the linear-dispersion approximation, $W_0$ shares a pole with $S_0$, i.e., at $1/\hat{n}_g$, whereas in the full theory $W_0$ does not increase as fast and remains finite, because the finite width of the absorption profile clamps the increase in mode width. Moreover, $M_0$ converges to a finite value, whereas it vanishes in the linear-dispersion approximation. Note the full theory better matches the data for both $W_0$ and $M_0$, although the measured value of $M_0$ is still consistently higher than the full theoretical prediction at all temperatures (by a constant shift of 0.078). This offset could be the result of a small systematic error in the determination of the modulation depth for the off-resonant mode.

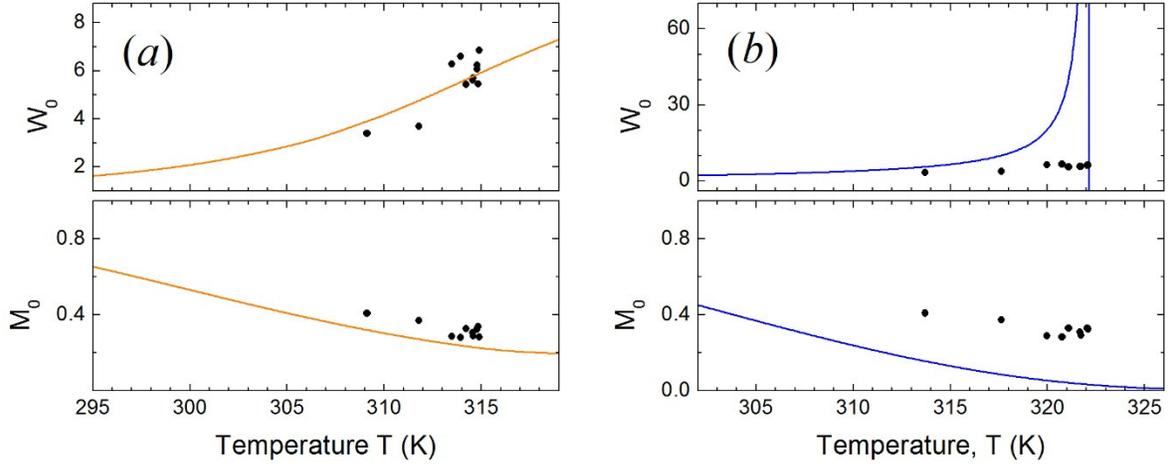

Fig. 5. (Color online) Factors that reduce sensitivity: on-resonance normalized mode width (top) and signal-to-noise (bottom) vs. vapor-cell temperature. For the data points, the temperatures were determined from the nonlinear least squares fits, whereas $W_0$ and $M_0$ were determined by direct measurement from cavity spectra. The data points are compared with the (*a*) full theory and (*b*) linear-dispersion approximation, using average values for $r_1$ and $a_1$. The temperature of the pole in $W$ is indicated by the vertical line in (*b*).

In Fig. 6(a) the QNL sensitivity enhancement obtained from Eq. (6) is plotted against the temperatures obtained from the model. As expected the enhancement is less than unity below the critical temperature, but then rises above unity closer to the critical temperature. Again, the data is consistently higher than the theoretical prediction owing to the offset in $M_0$. The largest value measured was 18.8 (with errors of +61 and −8.5). Note at low temperatures the error is roughly constant as it is determined primarily by the measurement of $W_0$ and $M_0$, but as the critical temperature is approached the error increases dramatically due to the increasing contribution from the scale factor measurement.

A comparison between the full theory and linear-dispersion approximation is shown in Fig. 6(b). Note in the linear-dispersion approximation there is no enhancement in QNL sensitivity, because $S_0$ and $W_0$ both increase as $1/\hat{n}_g$. Therefore, the scale-factor pole at 322.13K does not appear in the overall QNL enhancement. In the full theory, on the other hand, the pole in scale factor persists in the QNL sensitivity. Again, there are three reasons for this: (*i*) the scale factor increases faster than $1/\hat{n}_g$ due to mode reshaping, (*ii*) the increase in mode width is limited by the finite atomic absorption width / higher-order dispersion, and (*iii*) the decrease in signal-to-noise is limited because the cavity is not as far from critical coupling as it would be in the high-finesse linear-dispersion approximation.

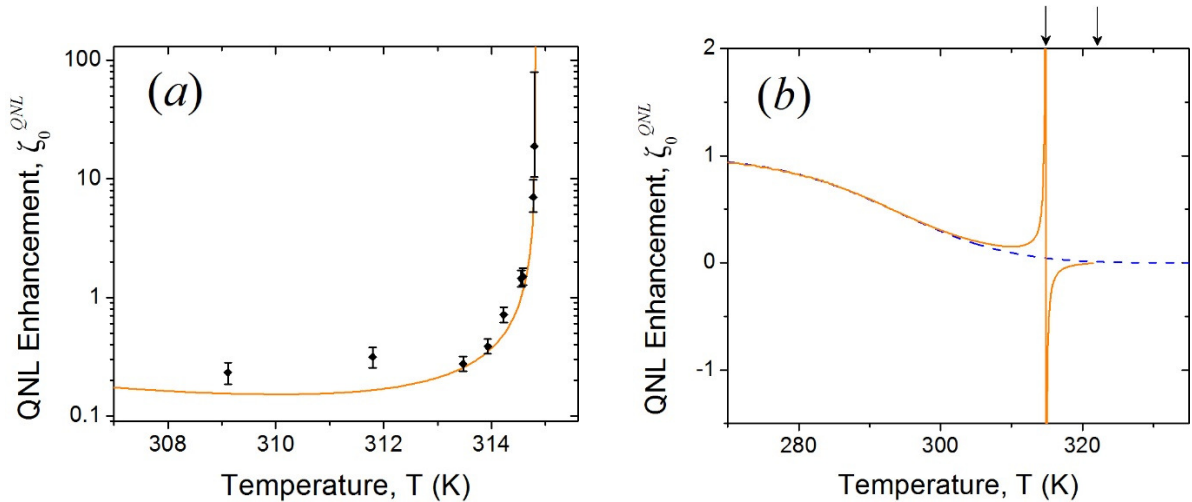

Fig. 6. (Color online) Predicted QNL sensitivity enhancement vs. vapor cell temperature. (a) Experimental results demonstrating that the enhancement rises above unity as the critical anomalous dispersion is approached. The solid curve represents the theoretical prediction. (b) Comparison of the full theory (solid curve) and linear-dispersion approximation (dashed curve). For each case, an arrow indicates the temperature where the scale-factor pole occurs (314.82K for the full theory and 322.13K in the linear approximation) corresponding to the critical anomalous dispersion. The pole does not show up in the QNL enhancement in the linear-dispersion approximation, but does appear in the full theory.

## 7. MONTE-CARLO SIMULATIONS

Recall that a crucial assumption in the derivation of Eq. (6), and therefore in the calculation of the QNL enhancement in the previous section, was that the effect of the change in mode shape on the frequency uncertainty of the mode could be neglected, even as the CAD condition is approached. We clearly see in Fig. 2, however, the modes becoming more flat-topped as they approach the atomic resonance. Therefore, in this section we drop the assumption behind Eq. (6), and compute the frequency errors for the dispersive and empty cavities through numerical Monte-Carlo simulations. This procedure provides a more direct indication of the QNL enhancement through use of Eq. (4), which takes account the change in mode shape near the CAD condition.

In this section it will be more straightforward to work with transmission, rather than reflection. We begin by assuming that the cavity transmission spectrum $T(\omega) = |\tau(\omega)|^2$ represents a probability distribution for the occurrence of photons at frequency $\omega$, and then draw random sample distributions from this underlying master distribution. An alternative approach would be to start with a master transmission curve and simply add Poissonian noise to it to generate the sample distribution [25], but we found this procedure yielded similar results and so do not present it here.

The central limit theorem states that the mean $\mu$ of a sample probability distribution will be normally distributed regardless of its particular shape. As a consequence, the standard deviation of the mean will vary with the number of samples $N$ according to $\sigma_\mu = \sigma / \sqrt{N}$, where $\sigma$ is the standard deviation of the original sample distribution. For unimodal distributions that are not skewed, the mean provides an estimate of the peak of the underlying distribution. It follows that, in the quantum noise limit, the frequency error associated with the determination of the transmission peak is

$$\delta\omega_{QNL} = \frac{\sigma}{SNR}, \tag{20}$$

where $SNR = \sqrt{N}$ is the signal-to-noise ratio at the detector, $N$ is the number of photons, $\sigma$ is the standard deviation of $T(\omega)$, and we have assumed for convenience that detector efficiency is unity. A requirement for Eq. (20) to be valid is that the *SNR* include all the photons in the sample distribution. This definition for the *SNR* is slightly different than in the discussion leading up to Eq. (9) in section 2, which is used to calculate *M* from our experiment, where only photons at the frequency peak were considered to contribute to the signal. We point out that the frequency range of the sample photon distribution depends on the particular experimental design (the frequencies of passive gyros, for example, are locked to the cavity resonances), and this distinction is negligible when the range is small. Moreover, while the absolute value of the *SNR* obtained from the two definitions can be quite different, the normalized quantity *M* is not significantly modified and varies similarly with cavity scale factor (in the limit of high finesse the dependency of *M* on scale factor is the same), irrespective of the choice of definition.

We have verified the validity of Eq. (20) for empty-cavity mode spectra over a large range of *SNR* by performing Monte-Carlo simulations to calculate $\sigma_\mu$ for a large number of random photon distributions. A practical difficulty arises because these spectra are not unimodal, and do not possess a standard deviation that converges, at infinite sample range. Spectra can be truncated to a single free spectral range such that Eq. (20) applies. Even so, the standard deviation will increase with sample range because cavity spectra are approximately Lorentzian at high finesse, and therefore always have significant probability density in their wings. It therefore makes sense in this case to use a range-independent parameter such as the full-width-at-half-maximum (FWHM) $\gamma$ to characterize the mode width, i.e.,

$$\delta\omega_{QNL} = C\frac{\gamma}{SNR}, \tag{21}$$

where the constant $C = \sigma/\gamma$ is dependent on range and cavity finesse. Note that the approximate relation of Eq. (5), relied upon extensively in the gyroscope literature, has been replaced by an exact relationship, i.e., Eq. (21). Notably, for high finesse cavities (such as gyroscopes) $C$ can differ significantly from unity.

An additional difficulty applies for cavities that contain a dispersive medium. In this case the proportionality constant in Eq. (21) also depends on the scale factor $S$, because the mode broadens and therefore takes up a larger portion of the data range. In this case, Eq. (6) can be rewritten as

$$\zeta^{QNL} = S \cdot \frac{\delta\omega_{QNL}^{(e)}}{\delta\omega_{QNL}} = K \frac{S \cdot M}{W}, \qquad (22)$$

where $K = C^{(e)}/C$. Moreover, because the modes can be skewed when the cavity mode is shifted slightly from the medium resonance frequency, the distribution mean does not accurately represent the mode peak. Eqs. (20)-(22) are not, in general, valid under these circumstances. Nevertheless, the peak can still be found by curve fitting, other averaging techniques, or simply by locating the mode of the distribution. This approach has the advantage that, to the extent they may involve less averaging, these alternate peak finding methods can be less dependent on data range. The distribution mode, for example, is independent of data range for sufficiently large $SNR$. In fact, one would expect the mode method to produce a limiting worst-case-scenario result as it uses the least amount of averaging, whereas the mean method represents the best-case limiting behavior, i.e., they represent cases of maximum and minimum amounts of averaging, respectively. Hence, the value of $\zeta^{QNL}$ that can be achieved in a given situation depends on the method used to identify the peak of the photon distribution, and how much averaging is involved in that method.

Therefore, we performed Monte-Carlo simulations using both these limiting methods. For each data point, the error $\delta\omega_{QNL}$ was obtained by taking the standard deviation of the mean or mode values obtained from 250 different photon distributions. The data range was limited to a single empty-cavity FSR, centered on the transmission peak. The same cavity parameters were used as in section 6, namely $a_1 = 0.674$, $r_1 = 0.946$, and FSR of 703.0 MHz. In addition, it was assumed that a second identical mirror with $r_2 = 0.946$ was used as an output coupler, so that cavity transmission could be monitored (rather than reflection as in section 6). The finesse in this case is 6.5, slightly lower than that used in our experiments in section 6. Photon distributions were produced by mapping a uniform random distribution onto the cavity transmission function through inverse transform sampling, i.e., by taking the squared-magnitude, integrating, and inverting Eq. (11). The maximum scale-factor enhancement was varied from $S_0 = 1$ to $S_0 = 344$ by changing temperature, and signal-to-noise was varied by changing the number of photons. In addition, as the scale factor increased, the photon number (and *SNR*) was reduced to account for attenuation of the mode as a result of medium absorption.

In Fig. 7, the ratio $K = C^{(e)}/C$ is plotted in transmission as the scale factor is varied. Recall that for the mean method, knowledge of $C$ and $K$ allows the frequency errors and QNL enhancement to be determined apriori by application of the central limit theorem (Eqs. (21) and (22)). Hence, for the top curve in Fig. 7, $K$ is calculated directly from the values of $\sigma$ and $\gamma$ for the master photon distribution, and is therefore independent of *SNR*. Note also that the value of $K$ asymptotes to a constant at high scale factor. This convergence of $K$ occurs because, as the scale factor increases, the broadening of the mode is eventually clamped by the finite absorption resonance width, i.e., by higher-order dispersion. At higher values of the finesse, $K$ converges to a larger value. For the bottom curves, on the other hand, the value of $K$ was determined aposteriori, i.e., by using Eq. (22) to back out the value of $K$ after computation of the frequency errors by the mode method. In this case, $K$ again approaches a constant value as the scale factor increases, but the particular constant value that it takes does depend on the *SNR*. As the *SNR* increases, however, the value of *K* eventually converges to a unique constant value below unity. Further explanation for this is given in the discussion below.

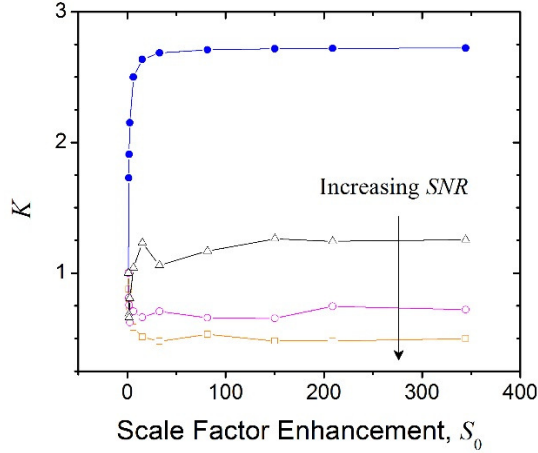

Fig. 7. (Color online) Top, closed symbols: The factor $K = C^{(e)}/C$, calculated directly from the master photon distribution converges to a constant value as the scale-factor enhancement $S_0$ increases. Bottom, open symbols: The factor *K* calculated from the frequency errors obtained from the mode method. The initial photon numbers (at $S_0 = 1$, $T = 210K$) are $10^3$, $10^4$ and $10^6$ (SNR values of 32, 100, and 1000), and decreases as $S_0$ increases. In the high scale-factor limit, *K* again becomes constant, but its value depends on *SNR*.

Fig. 8(a) shows the peak frequency error, $\delta\omega_{QNL}$, computed using the mean and mode methods discussed above, for the empty cavity ($S_0 = 1$) and dispersive cavities ($S_0 = 344$). The thick solid curve for the mean method is calculated from the RHS of Eq. (22), i.e., using the central limit theorem, and agrees with the simulation data. The peak frequency error for the two methods is identical when $SNR = 1$. Then, as the *SNR* increases, the error for the mean method decreases rapidly and monotonically, whereas the error for the mode method undergoes an initial increase before eventually decreasing. This initial increase occurs because there are not enough photons to produce more than one photon inside any given frequency interval. Subsequently, both methods reach a plateau, where further increases in *SNR* are of less benefit. Note that the peak frequency error for the mode method is considerably higher than for the mean method. This is not a general result, but occurs in this case because the finesse is low. At higher values of the finesse, we find comparable errors for the two methods. This is expected because as the finesse increases the mean and mode of the distribution approach one another.

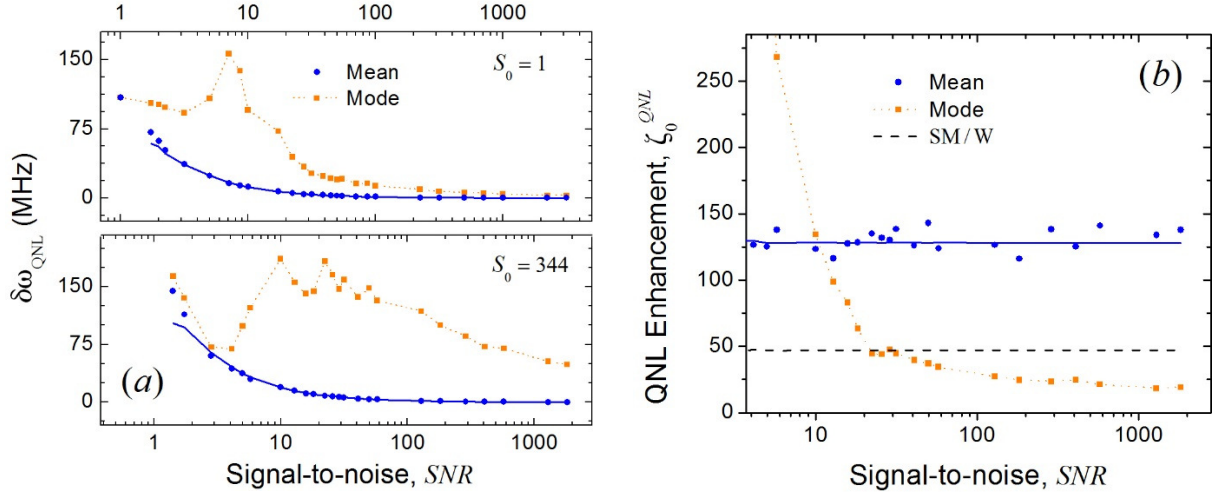

Fig. 8. (Color online) (*a*) QNL frequency error, $\delta\omega_{QNL}$, at constant scale factor, $S_0 = 1$ (top) and $S_0 = 344$ (bottom), calculated using the mean and mode methods. Note that for the same initial photon number, *SNR* is lower at the higher scale factor. Both methods eventually reach a plateau where further increases in *SNR* have little added benefit. (*b*) QNL enhancement, $\zeta_0^{QNL}$ for $S_0 = 344$. The dashed line indicates the result obtained by setting $K = 1$, i.e., the approximate result of Eq. (6). The thick solid curves for the mean method are calculated from the RHS of Eqs. (21) and (22), i.e., using the central limit theorem.

Fig. 8(b) shows the enhancement, $\zeta_0^{QNL}$, computed from the errors in Fig. 8(a) using both methods, at a fixed value of the scale factor ($S_0 = 344$), as *SNR* is varied. We have assumed, in computing $\zeta_0^{QNL}$, that the method used for finding the transmission peak is the same for the dispersive and empty cavities. The dashed line in the figure represents the QNL enhancement predicted by Eq. (6), which falls between the upper and lower limits obtained from the mean and mode methods, respectively. Whereas the mean method yields results that are independent of *SNR*, the mode method produces enhancements that depend on *SNR*, and this dependency varies with the scale factor. However, at sufficiently high *SNR*, we find that the enhancement asymptotes to a finite (nonzero) value. This convergence to a nonzero value is a result of not being in the linear-dispersion regime. Consider that in the linear regime the mode broadens to infinite width and assumes a flat-top shape as the critical anomalous dispersion is approached. For a flat-top shape the error determined from the mode method would be finite even at infinite *SNR*. On the other hand, for the empty cavity $(S_0 = 1)$ the error tends towards zero as the *SNR* becomes infinite because the mode shape is not flat-topped. Therefore, $\zeta_0^{QNL}$ would asymptote to zero at high *SNR*. In contrast, we find that $\zeta_0^{QNL}$ converges to a nonzero value, which reflects the fact that the linear-dispersion approximation is violated at high scale factor. Although not shown in the figure, we have confirmed that the *SNR* required for this convergence increases with cavity finesse.

The convergence to a finite QNL sensitivity enhancement is reiterated in Fig. 9(a), where the QNL enhancement $\zeta_0^{QNL}$, computed using the mode method, is plotted against the scale-factor enhancement, $S_0$. The mode method is not independent of *SNR*, rather it produces curves of $\zeta_0^{QNL}$ vs. $S_0$ whose slope is lower for higher initial *SNR* values. Nevertheless, as the *SNR* increases, the slope eventually converges to the asymptotic value, corresponding to the flat region in Fig. 8(b). Fig. 9(b) shows the QNL sensitivity enhancement computed using both methods. Again, the thick solid curve is calculated from the RHS of Eq. (22), for the largest initial photon number, and agrees well with the simulation data obtained using the mean method, independent of the *SNR*. For the mode method, we only include the results for the highest initial photon number $(N = 10^6)$, where the slope no longer changes with *SNR*.

Also plotted in Fig. 9(b) is the QNL enhancement calculated from Eq. (6), i.e., by setting $K = 1$ in Eq. (22), which again falls between the limits established by the two methods. Note at this value of *K*, the scale factor where the QNL enhancement crosses unity, i.e., where $\zeta_0^{QNL} > 1$, occurs at $S_0 = 6.4$. On the other hand, this same crossing point, determined from the experiment described in section 6, occurs at $S_0 = 25$ (see Fig. 4(b) and Fig. 6(b)). This difference is expected because, as pointed out in section 3, the scale factor in transmission is generally different from that in reflection owing to the differing amounts of reshaping that arise from the factor *F*. In addition, there is a slight difference in cavity finesse between the two cases due to the addition of the second output coupler.

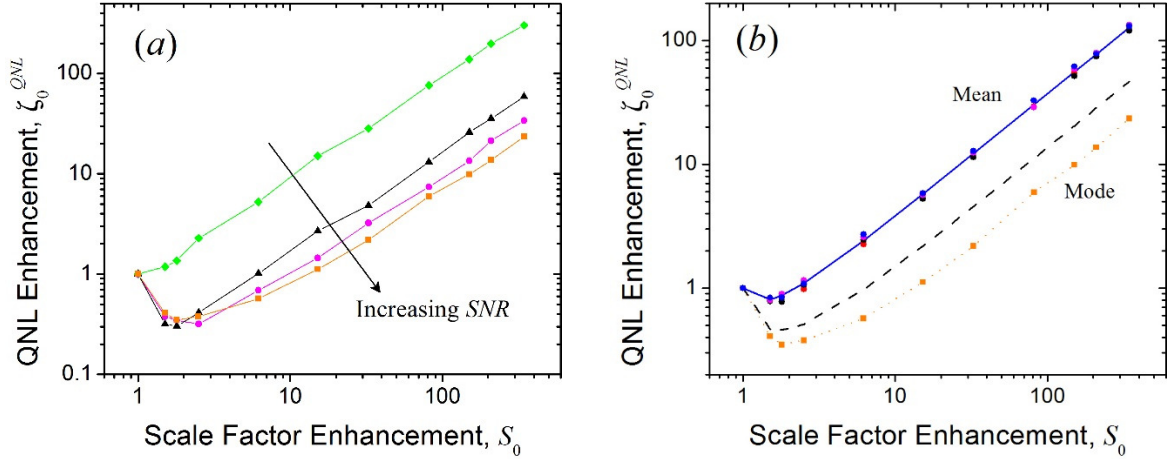

Fig. 9. (Color online) (a) QNL enhancement, $\zeta_0^{QNL}$, obtained from the mode method vs. scale-factor enhancement. At high scale factor, the curves are all linear. As the *SNR* increases the slope *K* decreases, eventually reaching the asymptotic value at high *SNR*. The initial ($S_0 = 1$) photon numbers were $N = 10^2$, $10^3$, $10^4$, and $10^6$ corresponding to *SNR* values between 10 and 1000. (b) QNL enhancement obtained from the mean (top, circles) and mode (bottom, squares) methods. For the mean method, different *SNR* values are represented by colored data points of varying color. For the mode method, only the asymptotic value is shown (initial photon number was $N = 10^6$). The solid curve (top) is calculated from the RHS of Eq. (22). The dashed line is the result obtained from Eq. (6).

Note, in Fig. 9(b), for both methods the enhancement initially decreases below unity, before eventually increasing proportionally with $S_0$. This is because, as discussed in the main text, $W_0$ and $M_0$ approach finite values at the critical temperature. Therefore, in this high scale-factor limit $(S_0 \gg 1)$, the mode method produces QNL enhancements that are also described by Eq. (22), but with a constant value of *K* that differs from that of the mean method. The values of *K* calculated from the mode method for different initial *SNR* values are presented in Fig. 7, where they are seen to reach constant values at high $S_0$. Moreover, while the particular constant value that *K* reaches depends on the initial *SNR*, a unique value is asymptotically approached at high *SNR*. Note that this procedure is lacking in comparison with application of the central limit theorem, which allows the frequency errors and QNL enhancement to be determined apriori by computing the constants

*C* and *K* directly from the standard deviation and FWHM of the master photon distribution. Nevertheless, that the ratio *K* takes a unique constant value in the limits of high scale factor and *SNR*, signifies that the mode method can be used to readily establish a lower bound for the QNL enhancement. In particular, once the value of *K* is established, knowledge of $S_0$, $M_0$, and $W_0$ is all that is required to characterize this lower bound. Finally, we note that in many experiments, e.g., in passive gyroscopes, the data range may be reduced far below one FSR. For such cases, the curves in Fig. 9(b) would tend to converge to the mode result, and the $K=1$ curve would fall outside the two bounds, revealing the inexactness of the approximation in Eq. (6).

## 8. SUMMARY AND CONCLUSION

Two distinct approaches were taken to investigate the possibility of obtaining a QNL sensitivity enhancement in the measurement of the mode frequencies of a passive cavity containing an anomalous dispersion medium. First, the scale-factor enhancement, normalized mode width, and signal-to-noise were simultaneously measured as the intracavity atomic medium is temperature tuned through the critical anomalous dispersion. Scale-factor enhancements larger than two orders of magnitude were obtained, translating to an order of magnitude enhancement in the predicted QNL measurement precision. A semi-empirical absorption model valid for low intracavity intensities was used to extract the temperatures and scale-factor enhancements. The accuracy of this model could be further improved by reducing systematic errors, i.e., through better design of the vapor cell oven and ensuring the proper empirical relationship between number density and temperature. Nevertheless, while these errors can shift the extracted temperatures (and, in turn, the critical temperature), they have negligible effect on the scale-factor enhancements. This is because, as we pointed out previously [6], the theoretical scale factor vs. temperature curve follows a $\sim 1/T$ phenomenological relationship, regardless of whether the full treatment or the linear-dispersion approximation is used, or whether adjustments in the critical temperature occur as a result of correcting these systematic errors. Our model is, therefore, sufficiently robust to extract scale-factor enhancements by fitting, even in the presence of systematic temperature errors (~1K). Moreover, the normalized mode width and signal-to-noise were determined directly from the cavity spectra by a model-independent process. Therefore, the enhancement in QNL precision, which is inferred from these quantities, is not significantly changed by these systematic errors.

On the other hand, the assumption that we used to calculate the QNL error, tacitly ignored the effect of the change in mode shape, which is dramatic near the critical anomalous dispersion condition. To account for this, a second approach was taken that bypasses this assumption by computing the frequency errors directly through numerical Monte-Carlo simulations using the theoretical cavity spectrum as a master photon probability distribution function. The results of this numerical procedure were found to depend on the particular averaging method involved in finding the distribution peak. Nonetheless, we confirmed that the same product of *S*, *W*, and *M* can be used to deduce the QNL enhancement, after multiplication by a constant factor *K* whose value is found to fall within a set of bounds (see Fig. 7).

Both approaches demonstrate the possibility of achieving an enhancement in measurement sensitivity in the QNL. To confirm the increase in precision, either in a passive ring gyroscope or linear configuration, the cavity would have to be held near QNL conditions, i.e., in a temperature-controlled vibration-free enclosure. This will be more difficult to achieve than for the equivalent empty cavity, in particular due to the temperature dependence of the atomic absorption [6]. It might be possible, however, to use a coupled-cavity, or other all-photonic design, rather than an atom-cavity configuration to reduce the temperature dependency. Such a confirmation would be an important proof-of-concept for the development of fast-light cavities and gyroscopes as a metrology tool.

## ACKNOWLEDGMENTS

This work was sponsored by the NASA Space Technology Mission Directorate Game Changing Development Office as well as the U.S. Army Aviation and Missile Research Development and Engineering Center (AMRDEC) Missile S&T Program.

20. D. D. Smith, H. Chang, L. Arissian, and J. C. Diels, "Dispersion enhanced laser gyroscope," Phys. Rev. A **78**, 053824 (2008).

21. G. A. Sanders, M. G. Prentiss, and S. Ezekiel, "Passive ring resonator method for sensitive inertial rotation measurements in geophysics and relativity," Opt. Lett. **6**, 569-571 (1981).

22. J. Gea-Banacloche, "Passive versus active interferometers: Why cavity losses make them equivalent," Phys. Rev. A **35**, 2518 (1987).

23. D. A. Steck, "Rubidium 87 D Line Data," available online at http://steck.us/alkalidata (revision 2.1, 1 September 2008)

24. S. J. Davis, W. T. Rawlins, K. L. Galbally-Kinney, and W. J. Kessler, "Spectroscopic investigations of Rb- and Cs- rare gas systems," in *High Energy/Average Power Lasers and Intense Beam Applications III*, SPIE Proceedings Vol. 7196, p. 71960G edited by S. J. Davis, M. C. Heaven, and J. T. Schriempf (SPIE, Bellingham, 2009).

25. W. M. van Spengen, "The accuracy of parameter estimation by curve fitting in the presence of noise," J. Appl. Phys. **111**, 054908 (2012).

36